\documentclass[
  twoside,
  twocolumn,
  prb, 
  aps,
  floatfix,
  preprintnumbers, 
  floatfix,    
  superscriptaddress,
  showpacs
]{revtex4-1} 

\usepackage{amsmath}
\usepackage{amssymb}
\usepackage{color}
\usepackage{graphicx}
\usepackage{times}

\newcommand{\Cd}[0]{\text{Cd}}
\newcommand{\Zn}[0]{\text{Zn}}
\newcommand{\Te}[0]{\text{Te}}

\newcommand{\CxZxT}[0]{\text{Cd}_{1-x}\text{Zn}_x\text{Te}}
\newcommand{\CZT}[0]{\text{CZT}}
\newcommand{\bulk}[0]{\text{bulk}}

\newcommand{\VBM}[0]{\text{VBM}}

\newcommand{\meV}[0]{\text{meV}}

\newcommand{\braket}[3]{\bigl{<}#1\big|#2\big|#3\bigr{>}}

\newcommand{\sect}[1]{Sect.~\ref{#1}}
\newcommand{\fig}[1]{Fig.~\ref{#1}}
\newcommand{\eq}[1]{(\ref{#1})}

\newlength{\myhgt}
\usepackage[usenames,dvipsnames]{xcolor}
\newcommand{\onlineref}[1]{Ref.~\onlinecite{#1}}

\renewcommand{\epsilon}{\ensuremath\varepsilon}
 
\begin{document}

 \preprint{}

 \title{
   Contributions of point defects, chemical disorder, \\ and thermal vibrations to electronic properties of $\text{Cd}_{1-x}\text{Zn}_x\text{Te}$ alloys
 }

\author{Daniel \AA{}berg}
\affiliation{
  Physical and Life Science Directorate, Lawrence Livermore National Laboratory, California 94550, USA
}

\author{Paul Erhart}
\affiliation{
 Physical and Life Science Directorate, Lawrence Livermore National Laboratory, California 94550, USA
}
\affiliation{
  Chalmers University of Technology, Department of Applied Physics, S-412 96 Gothenburg, Sweden
}

\author{Vincenzo Lordi}
\affiliation{
 Physical and Life Science Directorate, Lawrence Livermore National Laboratory, California 94550, USA
}

 \begin{abstract}
We present a first principles study based on density functional theory of thermodynamic and electronic properties of the most important intrinsic defects in the semiconductor alloy $\CxZxT$ with $x<0.13$. The alloy is represented by a set of supercells with disorder on the Cd/Zn sublattice. Defect formation energies as well as electronic and optical transition levels are analyzed as a function of composition. We show that defect formation energies increase with Zn content with the exception of the neutral Te vacancy. This behavior is qualitatively similar to but quantitatively rather different from the effect of volumetric strain on defect properties in pure CdTe. Finally, the relative carrier scattering strengths of point defects, alloy disorder, and phonons are obtained. It is demonstrated that for realistic defect concentrations carrier mobilities are limited by phonon scattering for temperature above approximately 150\,K.
 \end{abstract}
 

\maketitle

\section{Introduction}

Cadmium telluride (CdTe) has found applications in a wide range of areas including but not limited to solar cells,\cite{BriFer93} radiation detectors,\cite{Kno10} and electro-optical modulators.\cite{TriSif09}
CdTe is a prime candidate material for room-temperature radiation detectors thanks to a number of reasons:
(1) Its band gap lies within the range that maximizes the signal-to-noise ratio by balancing the numbers of photogenerated carriers {\it vs} thermally generated carriers.
(2) 
High purity CdTe crystals exhibit a long carrier drift length, characterized by the product of carrier mobility $\mu$ and carrier trapping lifetime $\tau$.\cite{DelAbbCar09, Owe06, RedPanPan09, SchTonYoo01} A large $\mu \tau$ value ensures efficient charge collection and a linear response to the energy of  incident radiation. 
(3) Finally, these samples possess also high resistivity $\rho$, which is required to keep leakage currents at a minimum.

Further improvements can be accomplished by addition of Zn to obtain $\CxZxT$ (CZT) alloys. The band gap gradually increases with Zn concentration from 1.5 to approximately 1.6\,eV at $x\approx0.1$, which leads to a further reduction of thermal noise and raises the resistivity up to $8\times10^{10}$ $\Omega$\,cm. \cite{DelAbbCar09, SchTonYoo01, MelSigVor99}

Zn alloying has empirically been found to have additional benefits such as oxygen gettering, increased hardness, and reduced degradation of detector performance over time from build-up of internal fields under high radiation flux (``polarization'' effect). \cite{DelAbbCar09} These features have allowed the fabrication of CZT gamma detectors with excellent energy resolution better than 1\% at 662~keV. \cite{ZhaZhoXu04,AwaCheMac08} The large scale deployment of CZT detectors is, however, hampered by high material costs. This results from the difficulty to grow large uniform single crystals, which requires cutting  the ingot into individual crystallites, followed by testing and selection (``crystal harvesting''). A variety of  defects---from point defects to dislocations and secondary phases---limit the performance of the individual crystals. Understanding and learning to control the different contributions is therefore key to improving the manufacturing yield and driving down cost.

Defects affect resistivity, charge carrier mobilities, and carrier lifetimes in different ways. The resistivity for example is largely determined by the concentrations of charged defects, which are coupled by the charge neutrality condition and can be obtained from the defect formation energies. \cite{ErhAlb08,ErhAbeStu10} Charge carrier mobilities in turn are limited by scattering at lattice perturbations and are therefore sensitive not only to point defects and dislocations but also thermal vibrations. Finally, lifetimes are shortened by carrier trapping, which occurs preferentially at defects that induce deep levels in the band gap leading to carrier recombination via the Shockley-Read-Hall mechanism. \cite{ShoRea52}
In CdTe and CZT, generally lifetime-limiting defects remain the biggest challenge to maximizing device performance.

These effects are modified in the case of alloys. The chemical disorder introduces a distribution of defect formation energies and electronic transition levels, and provides additional scattering channels.
In this work we compute the formation energies and transition levels of native defects in CZT and study their dependence on alloy composition as well the possible appearance of new defect levels in the gap. 
Furthermore, we determine the relative importance of point defects, phonons, and alloying for carrier mobilities. This assessment is important for developing material optimization strategies. For example intentional doping and annealing can be used to modify the defect populations with little impact on phonon properties.

There are approaches at different levels of sophistication for modeling the thermodynamic properties of an alloy including for example the virtual crystal approximation (VCA), \cite{Nor31, Nor31a} the coherent potential approximation, \cite{Sov67} special quasirandom structures, \cite{ZunWeiFer90,*WeiFerBer90} and alloy cluster expansions. \cite{LakFerFoy92} 
Here, in order to obtain a simple unified approach that allows for the study of both the distribution of defect-related thermodynamic quantities and scattering rates, we model the alloy disorder by using a set of supercells with Zn atoms randomly distributed over Cd sites. This approach has  been  applied successfully by Carvalho {\it et al.} to study formation energies of cation-site intrinsic defects in CZT. \cite{CarTagObe10} In the following we employ density functional theory to demonstrate from first principles that ({\em i}) point defect formation energies increase or stay approximately constant as a function of Zn concentration, ({\em ii}) optical transition levels are virtually unchanged upon alloying, and ({\em iii}) mobilities are limited by phonon scattering at ambient conditions. 
The study focuses on Zn concentrations up to 15\%, which is the most relevant range for applications in radiation detection.

\section{Methodology}
\label{sec:methodology}
\subsection{Point defect thermodynamics}

The temperature dependence of the equilibrium defect concentration obeys the following relation \cite{AllLid03} 
\begin{align}    
  c &= c_0 \exp \left( -\frac{\Delta G_f}{k_B T} \right),
  \label{eq:concentration}
\end{align}
where $c_0$ denotes the concentration of possible defect sites and $\Delta G_f$ is the Gibbs free energy of defect formation, which we approximate by the formation energy $\Delta E_f$, assuming that vibrational entropy and pressure-volume term are small.\cite{AbeErhWil08} The formation energy for a defect in charge state $q$ is given by\cite{QiaMarCha88, ZhaNor91}
\begin{align}
  \Delta E_f
  &= E_\text{def} - E_\text{ideal}
 - \sum_i \Delta n_i \mu_i + q \left( E_{\VBM} + \mu_e \right)
  \label{eq:eform}
\end{align}
where $E_\text{def}$ and $E_\text{ideal}$  are the energies of the system with and without the defect, respectively, $\Delta n_i$ denotes the number of atoms of type $i$ and chemical potential $\mu_i$ added to the system, and $\mu_e$ is the electron chemical potential measured with respect to the valence band maximum (VBM), $E_\VBM$.

Neglecting entropic contributions , the heat of formation for the CZT alloy is given by
\begin{align}
  \Delta H_f(\CZT,x) & = (1-x) \mu_\Cd + x \mu_\Zn + \mu_\Te  \nonumber \\
  & - (1-x) \mu_\Cd ^{\bulk} - x\mu_\Zn^{\bulk} -  \mu_\Te^{\bulk}. 
\end{align}
Defining $\mu_i=\mu_i^{\bulk}+\Delta \mu_i$, where $\mu_i^{\bulk}$ is the chemical potential of component $i$ in its reference state, we obtain the relation
\begin{align}
  \Delta H_f(\CZT,x) = (1-x) \Delta \mu_\Cd + x \Delta \mu_\Zn + \Delta \mu_\Te.
\end{align} 
Since $\Delta \mu_i$ has to be non-positive to avoid forming bulk material $i$, the range of the chemical potentials is constrained by
\begin{align}
  0 \leq \Delta \mu_i \leq  \Delta H_f(\CZT,x)
\end{align}
We then define the cation-rich limit as $(1-x) \Delta \mu_\Cd + x \Delta \mu_\Zn=0$ and the Te-rich limit by $\Delta \mu_\Te=0$ (see Ref. \onlinecite{RamFurBec02}).
Given the formation energies at $\mu_e=0$ for two different charge states, $q_1$ and $q_2$, we compute the equilibrium electronic transition level as
\begin{align}
  E^{el}_{tr}(q_1\rightarrow q_2) &= \frac{\Delta E_f(q_2)-\Delta E_f(q_1)}{q_1-q_2}
\end{align} 
 
\subsection{Theoretical carrier mobilities} 

In a quasi-classical picture, the motion of free carriers subjected to a weak electric field is governed by the Boltzmann transport equation.\cite{mahan2000} In particular, the electrical conductivity and carrier mobilities are related to scattering processes at the microscopic scale.\cite{YuCar10} In an alloy system, three major carrier scattering channels are active, namely those involving phonons, the inherent randomness of the alloy itself, and defects.\cite{YuCar10, Nor31, Nor31a, MurFah06, JoyMurFah07}  The relative importance of these scattering channels can be assessed from first principles using Fermi's Golden rule
\begin{align}
  R_{i\rightarrow f} = \frac{2\pi}{\hbar} 
  \left |
  \braket{f}{\Delta V}{i}
  \right |^2 \delta(\epsilon_f-\epsilon_i).
  \label{eq:scatstr}
\end{align}
Here $i$ and $f$ denote initial and final states, respectively, and $\Delta V$ is the relevant scattering potential. To compare the relative importance of each scattering channel, each contribution should be treated at the same level of approximation. Although highly accurate first-principles approaches have been developed to tackle each channel separately,\cite{MurFah06,JoyMurFah07,LorErhAbe10} the computational cost associated with the combinatorial explosion of point defects in an alloy currently hinders further exploration. Therefore, more cost-effective approaches are desirable. In the following we present an approximate, but comprehensive, approach that forms the basis for our analysis.

\paragraph{Defect scattering} 
For point defect scattering, we begin with Fermi's golden rule as shown in \eq{eq:scatstr}. 
The orthogonality of the initial and final states ensures that only gradients of $\Delta V$ contribute to the integral. Thus, we have earlier defined a simpler, but cruder, measure termed the \textit{relative scattering strength} by\cite{LorErhAbe10}
\begin{align}
M=
 \int d\mathbf{r} \left| \nabla_\mathbf{r} (\Delta V)\right |
.\label{eq:relscatstr}
\end{align}
Note that this quantity is independent of initial and final state wave functions. Despite its simple form, we have shown that $M^2$ from Eq. \eq{eq:relscatstr} is approximately proportional to the more computationally expensive Brillouin-zone average over all scattering rates given by Eq.~\eqref{eq:scatstr}.\cite{LorErhAbe10} 

\paragraph{Alloy scattering}
To quantify scattering due to alloy disorder, we follow a similar procedure as for defect scattering.
For a given Zn concentration $x$ we calculate the relative scattering strength using the scattering potential $\Delta V$ given by 

\begin{align}
  \Delta V_\text{alloy} &= V_\text{alloy} - (1-x) V_\text{CdTe} - x V_\text{ZnTe},
\end{align}
where all terms are computed for the same supercell size and lattice constant. We then average over an ensemble of random alloy configurations for each $x$. Substitution of $(1-x) V_\text{CdTe} + x V_\text{ZnTe}$ by the potential resulting from a VCA calculation at the same concentration did not change the results.

\paragraph{Phonon scattering} 
Here we  only consider phonon scattering in  pure CdTe  and assume the results to  transfer to the CZT alloy. This is justified since the vibrational frequencies and phonon polarization vectors do not change much for small concentrations of Zn in CdTe. Again following the perturbative approach as above, we compute the relative scattering strength of a thermal population of phonon modes by taking the difference between a thermally excited atomic configuration and the ideal zero temperature configuration. We perform the ensemble average over phonon modes by averaging over different atomic configurations representing thermally populated phonons. The configuration space was sampled in two different ways to separate harmonic and anharmonic contributions.
In the first approach, cells were generated by superimposing harmonic phonons modes with random phases and amplitudes chosen to yield the correct mean square displacement for the temperature of interest. In the second approach, anharmonicity was explicitly accounted for by sampling decorrelated configurations from {\em ab initio} molecular dynamics trajectories.

\subsection{Computational details}
\label{sec:compdet}

Density functional theory (DFT) calculations were carried out in the local density approximation (LDA) using the Vienna ab-initio simulation package (\textsc{vasp}) \cite{KreHaf93, *KreHaf94, *KreFur96a, *KreFur96b} and ultrasoft pseudopotentials. \cite{Van90} We employed supercells of 216 atoms sampled in reciprocal space with 2$\times$2$\times$2 $k$-point grids generated using the Monkhorst-Pack scheme. \cite{MonPac76} The plane wave energy cutoff was set to 277\,eV and Gaussian smearing with a width of 0.1\,eV was used to determine the occupation numbers. For a given Zn concentration we created 20 replicas of the original ideal or defective cell and randomly substituted Zn for Cd. The cells were subsequently relaxed until ionic forces were less than 30\,\meV/\AA. The lattice parameters were taken from VCA calculations.\footnote{The VCA for total energies, forces, and stress was implemented into \textsc{vasp} using the approach of Bellaiche and Vanderbilt. \cite{BelVan00} Note that this implementation is separate from the current official release.} A homogeneous background charge was added for calculations including a charged defect  to ensure charge neutrality of the entire cell. The monopole-monopole correction of Makov and Payne was applied to correct for the spurious interaction of charged defects.\cite{MakPay95} No band gap or potential alignment corrections were performed.
 
We note that equilibrium transition levels $E^{el}_{tr}$ are distinct from optical transition levels $E^{opt}_{tr}$ as the former determine the equilibrium electron chemical potential at which a change in the charge state occurs whereas the latter ones correspond to optical (``vertical'') transitions without the involvement of lattice relaxations.
Neglecting excitonic or multiplet effects we can in principle calculate optical transition levels from total energy differences of a defect in charge state $q$ and $q\pm1$ in the same geometry. Thus, we need an additional separate calculation of the total energy for each relaxed defect geometry. Due to the very large number of defect configurations we here instead make the additional approximation that $E^{opt}_{tr}$ may be calculated by Kohn-Sham eigenvalue differences for charge state $q$. We note that this corresponds to the neglect of changes in the DFT double counting term (see Ref. \onlinecite{AbeErhWil08}).

In this work, we focus on the most important defects as determined by earlier studies.\cite{WeiZha02,ErhAbeLor12} These are the Cd and Te vacancies, Te on Cd antisite (in local $T_d$ and $C_{3v}$ symmetries), Te-tetrahedrally coordinated Cd interstitials, and the most important Te-Te and Cd-Te dumbbell split interstitials. We refer to \onlineref{ErhAbeLor12} for details of the respective geometries. 

\section{Results and Discussions}
\label{sec:results}
\subsection{Ideal \CZT\ alloy}

\begin{figure}
 \includegraphics[width=\columnwidth]{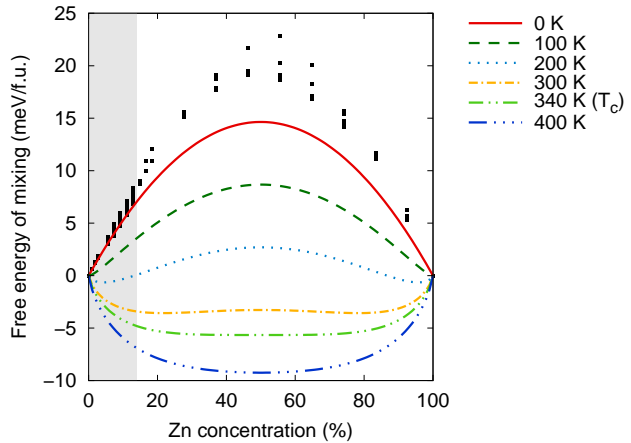}
 \caption{
   (Color online) Free energy of mixing of CZT as a function of Zn concentration and temperature. Symbols denote the enthalpies of mixing calculated using the total energies from the supercell approach. The red solid line is the result of a quadratic polynomial fit to the lowest energy configurations within the gray area ($\leq$ 13\%). The remaining lines represent the free energy of mixing up to 400~K, as generated by adding the entropy contribution for a random solution.
}
  \label{fig:mixenthalpy}
\end{figure}

The adequacy of the supercell approach, outlined in Sect. \ref{sec:compdet} to mimic the behavior of the ideal alloy was assessed by a study of its thermodynamic properties below the melting temperature. Firstly, the zero Kelvin enthalpy of mixing was obtained by fitting a second order polynomial to the lowest energy configurations for Zn concentrations up to 13\%, as displayed in \fig{fig:mixenthalpy}. This corresponds to the enthalpy of mixing for pairwise interaction energies and a random distribution of atoms on the sub-lattices.\cite{CheShe95} Thus, the system is completely immiscible at absolute zero. The critical temperature ($T_c$) was then approximated by assuming a completely random solution, yielding $T_c = 340$~K. In contrast, the VCA (results not shown) erroneously predicts a miscible solution at all temperatures.

The literature is unfortunately riddled with disparate estimates of $T_c$. Model calculations based on the CALPHAD approach \cite{MarDruTri92} give 701~K, while cluster expansion calculations based on first principles energies \cite{WeiFerZun90} predict 605\,K. Experimental estimates also vary, with one careful study reporting $T_c = 435$~K from a fit to extended x-ray absorption fine-structure characterization of phase separating alloys, but also $T_c = 340$~K by extracting parameters from the liquid--solid phase diagram.\cite{MotBalLet85}
Earlier experimental studies had reported and accepted higher values for $T_c$, but it is pointed out that slow diffusion at low temperatures makes the determination difficult by observation of phase separation, which might grossly overestimate the binodal stability temperature.

 \begin{figure} 
  \includegraphics[width=\columnwidth]{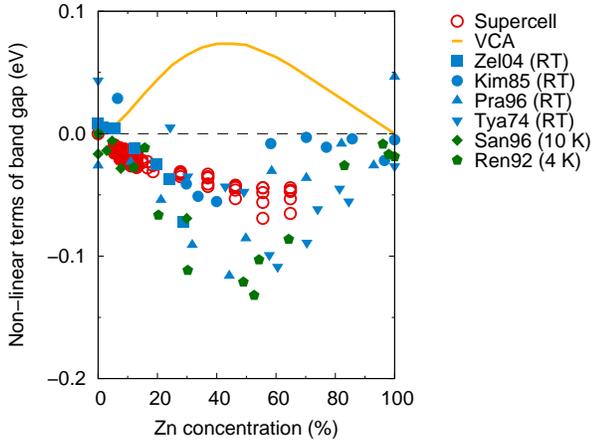}
  \caption{
    (Color online) Deviation from linear dependence of band gap as a function of Zn concentration. The solid yellow line and the red open circles denote the VCA and supercell results, respectively. The calculated values are shifted with a linear function to coincide with the room-temperature experimental band gap of CdTe and ZnTe at $x=0$ and $x=1$, respectively. Experimental values recorded at room-temperature are shown with  blue symbols, see Refs. \onlinecite{ZelMenBec04,KimMenSco85,PraHusRed96,TyaSniBon74,ChaSarCha91}. Data recorded below 10\,K are shown with green symbols, see Refs. \onlinecite{SanNavRam96} and \onlinecite{RenJon92}. The latter have been shifted by a constant to coincide with the room-temperature data.
  }
  \label{fig:bandgap} 
\end{figure}  

In addition, we calculated the variation in band gap as a function of Zn concentration and compared to experimental data, as shown in \fig{fig:bandgap}. The randomized supercell approach accurately reproduces the experimental trend, while the VCA exhibits incorrect bowing. In fact, even the sign of the bowing is incorrect with VCA over almost the entire range.

\subsection{Formation energies and electronic transition levels}
\label{sec:eform}
\begin{figure}
 \includegraphics[width=\columnwidth]{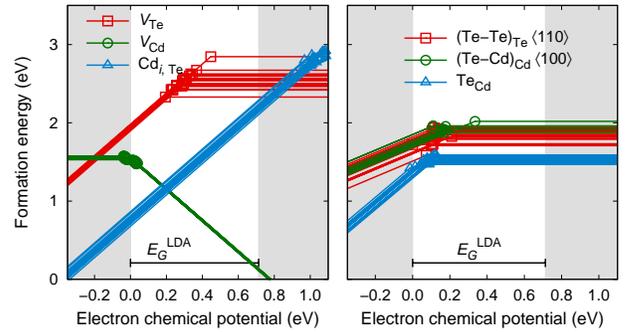}
  \caption{(Color online) Dependence of the formation energies on the electron chemical potential under Te-rich conditions for the most important intrinsic defects in $\CxZxT$ ($x=0.08$). Each line corresponds to a separate random Zn configuration. The electronic (charge state) transition energies are denoted by the symbols.}
  \label{fig:eformCZT_10}
\end{figure}

\begin{figure*}
 \includegraphics[width=\columnwidth]{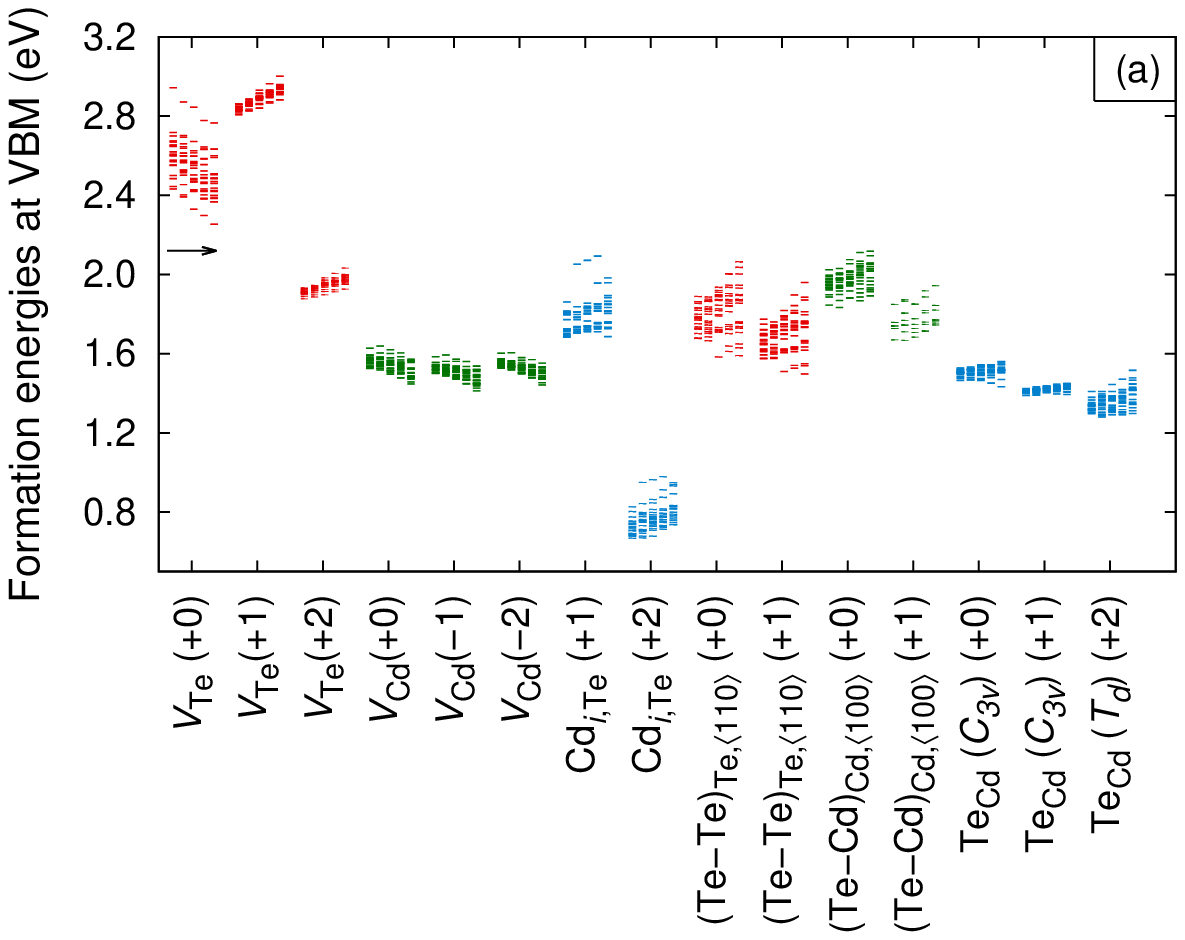} 
 \includegraphics[width=\columnwidth]{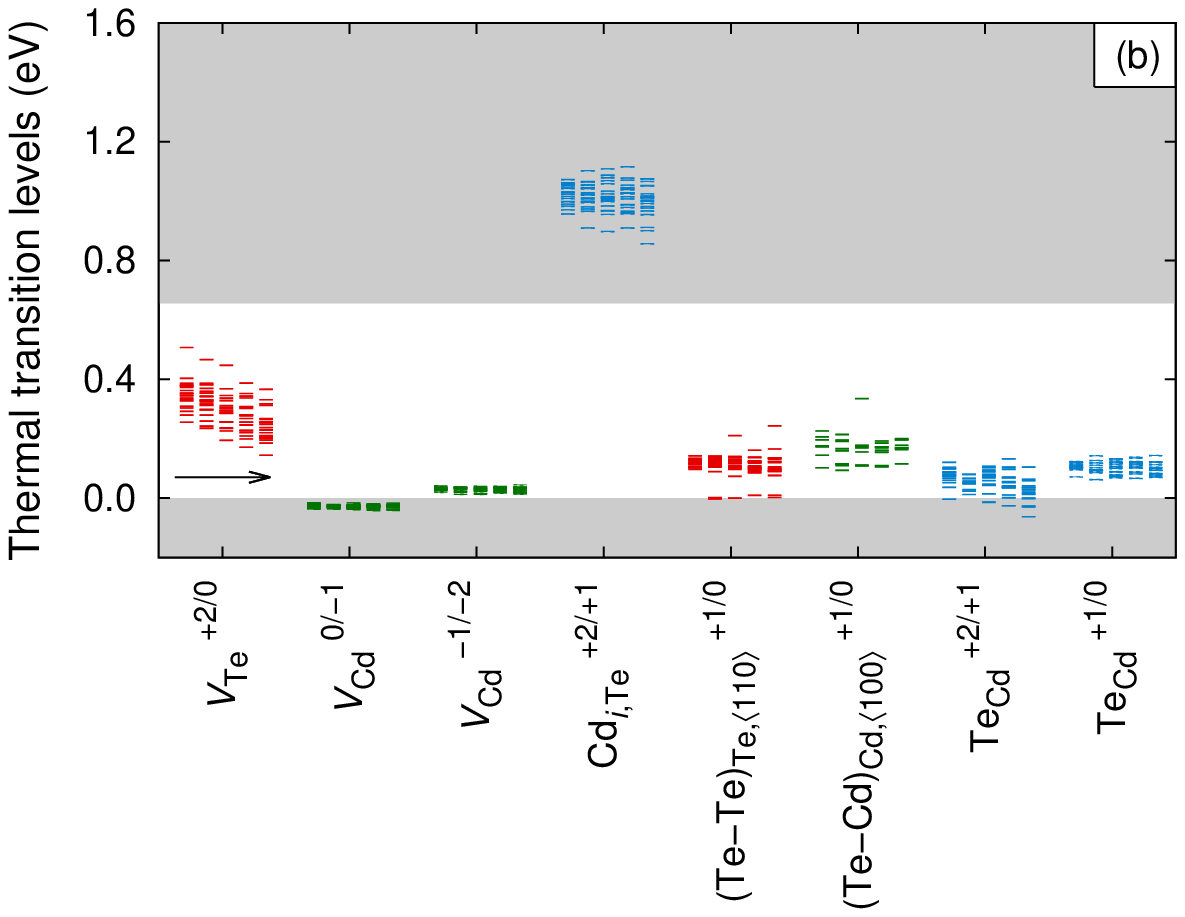} 
  \caption{
    (Color online) (a) For each defect and charge state, the formation energy at the VBM for Te-rich conditions is plotted for the set of configurations for five different concentrations ranging from 5.5 to 13.0\% (left to right, indicated by arrow).
    (b) For each defect, the electronic transition levels are plotted for the set of configurations for five different concentrations ranging from 5.5 to 13.0\% (left to right, indicated by arrow).
  }
  \label{fig:eformCZT}
\end{figure*}

We display in \fig{fig:eformCZT_10} the formation energies (lines) and electronic transition energies (symbols) as a function of the electron chemical potential under Te-rich conditions for the specific case of $x=0.08$.  To summarize, the most dominant defects are the double acceptor $V_{\Cd}^{2-}$, double donor $\Cd_{i,\Te}^{2+}$, and the electrically neutral $\Te_\Cd^0$. In fact, the relative ordering in terms of formation energies and electronic transition levels for the different defects is unchanged with respect to CdTe.\cite{ErhAbeLor12} 
However,  in the case of \CZT\ a few defects, most notably $V_\Te$ and $\Cd_{i,\Te}$, display a strong variation of formation energies and electronic transition levels between different alloy configurations (supercells). This effect is highlighted in \fig{fig:eformCZT}, where $\Delta E_f$ and $E^{el}_{tr}$ are plotted separately for each configuration as a function of Zn concentration.

In principle, strong variations in the formation energies are a tell-tale sign of defects being sensitive to local structure variations. The impact of Zn composition and alloy disorder on formation energies will now be discussed the cases of $V_\Te$ and $\Cd_{i,\Te}$.

 \begin{figure}
 \includegraphics[width=\columnwidth]{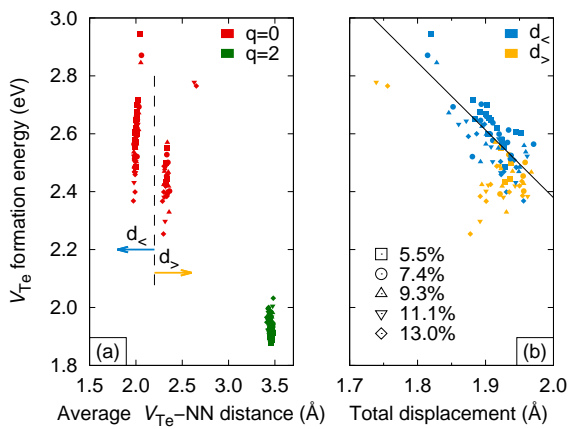}
 \caption{
   (Color online) (a) $V_\Te$ formation energy at the VBM as a function of average $V_\Te$-nearest neighbor distance. The Zn concentration is indicated by the different symbols. Red and green symbols correspond to charge states $q=0$ and $q=2$, respectively. Dashed line indicate separation of configurations with small (d$_<$) and large(d$_>$) $V_\Te$-NN distances. (b) $V_\Te$ formation energy for $q=0$ at the VBM as a function of displacements of all atoms with respect to the defect free system. Again, symbols denote Zn concentration and colors indicate small (d$_<$) and large(d$_>$) $V_\Te$-NN distances. The line is drawn to guide the eye.
 }
  \label{fig:geomVTe}
\end{figure} 

\subsubsection{Tellurium vacancy}
In contrast to the anion vacancies in III-V semiconductors, \cite{BerPan78,TalTin82,Pus89,AbeErhWil08} the $V_\Te$ defect is not subject to a Jahn-Teller distortion. \cite{LanOstWol01} Rather, in an unrelaxed supercell containing a single neutral Te vacancy, the $a_1$ defect level resides inside the band gap and is fully occupied, whereas the triply degenerate $t_2$ level is completely empty. Thus no energy can be gained by breaking the $T_d$ symmetry. When allowing the ions to relax, the neighboring Cd ions move symmetrically inward by a distance of 2.0~\AA\ toward the vacant site for $q=0$ and outward by 3.5~\AA\ for $q=2$. The inward relaxation for $q=0$ causes the $a_1$ defect level to lower its energy and hybridize with valence band states. In \CZT\ the $V_\Te^{2+}$ geometry is virtually unchanged with respect to the defective CZT, whereas the geometry of the $q=0$ state displays drastic unsymmetric variations. In particular, the nearest neighbor (NN) Zn atoms tend to not relax inwards as the Cd NNs do. We however emphasize that in some configurations Cd NN also display this behavior. As shown \fig{fig:geomVTe}a, this manifests itself as two two distinct average $V_\Te$--NN distances (d$_<$ and d$_>$). Furthermore, the unsymmetric configurations (d$_>$), which in most cases involves a Zn NN, tend to have lower formation energies. Note that this effect is not observed for charge state $q=2$. However, as the overlap of the formation energies for the two distinct groupings is appreciable, the large variation cannot be correlated to the number of NN Zn neighbors or local geometry alone.

Figure~\ref{fig:geomVTe}b displays the variation of formation energies with respect mean square displacement $\sum_i\Delta r_i$ of all atoms. It can be seen that the formation energies decrease with increasing amount of relaxation. The reduction is roughly linear for the geometries that maintain the \emph{local} $T_d$ symmetry. Thus we conclude that the variation in formation energies is due to atomic relaxation outside the nearest neighbor shell. This is consistent with the earlier observation that the occupied $a_1$ defect state hybridizes with extended valence band states.

\subsubsection{Cadmium interstitial}
\begin{figure}
 \includegraphics[width=\columnwidth]{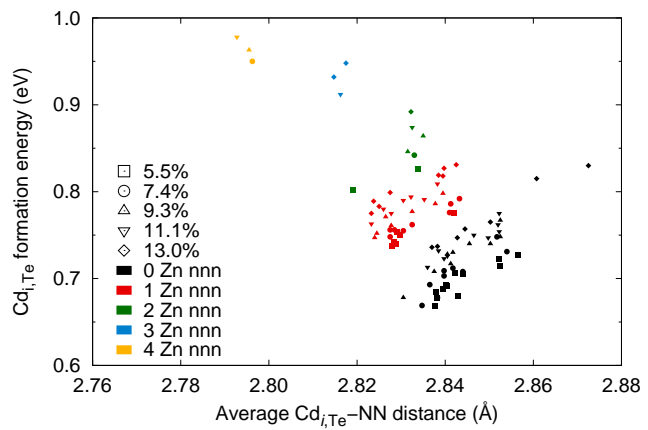}
  \caption{(Color online) Formation energy at the VBM as a function of average Cd$_{i,\Te}$-nearest neighbor distance. The symbols and colors indicate the Zn concentration and the number of Zn atoms in the next-nearest neighbor shell.
  }
  \label{fig:geomCdi2}
\end{figure}

The analysis of the Cd$_{i,\Te}$ geometries is  more straightforward. In CdTe, the Cd$_{i,\Te}^{2+}$ is tetrahedrally coordinated to four Te, each at a distance of 2.85 \AA. By random substitution of Cd by Zn, the formation energies spread to a range of about 0.35~eV. In \fig{fig:geomCdi2} we show the formation energies as a function of average NN distance and number of next-NN Zn atoms. One finds the average formation energy to generally increase with Zn concentration (see \fig{fig:eformCZT}). Furthermore, the large variation of formation energies is found to be strongly dependent on the number of next-NN Zn atoms which also correlates with the average NN distance.  
 
\subsubsection{Alloying {\it vs} strain}
\begin{figure*}
 \includegraphics[clip=true,trim=40 40 55 0,width=0.25\paperwidth ]{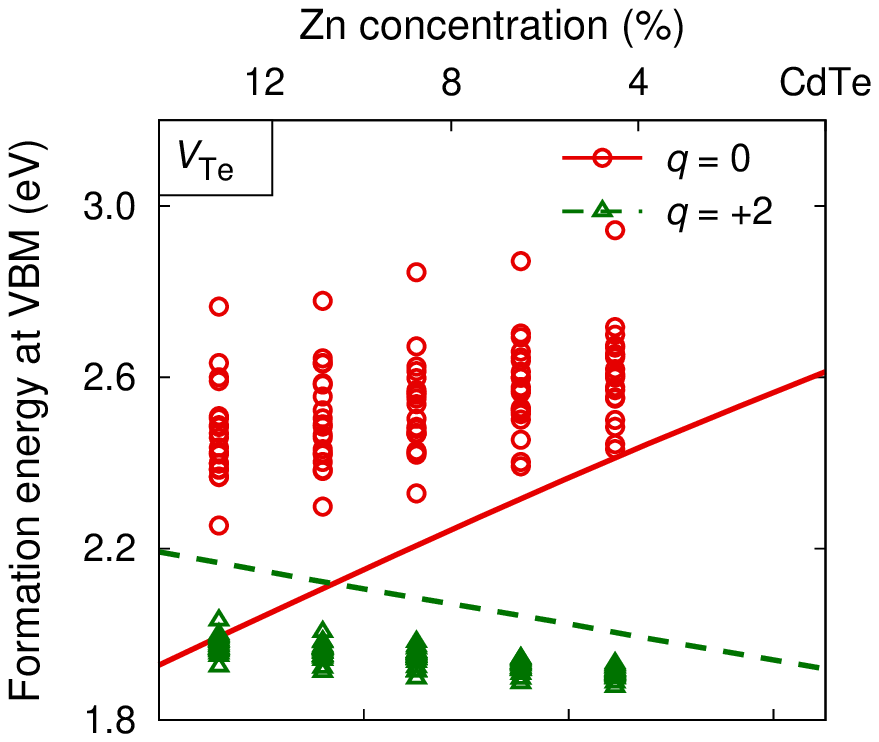} \nolinebreak
 \includegraphics[clip=true,trim=40 40 55 0,width=0.25\paperwidth ]{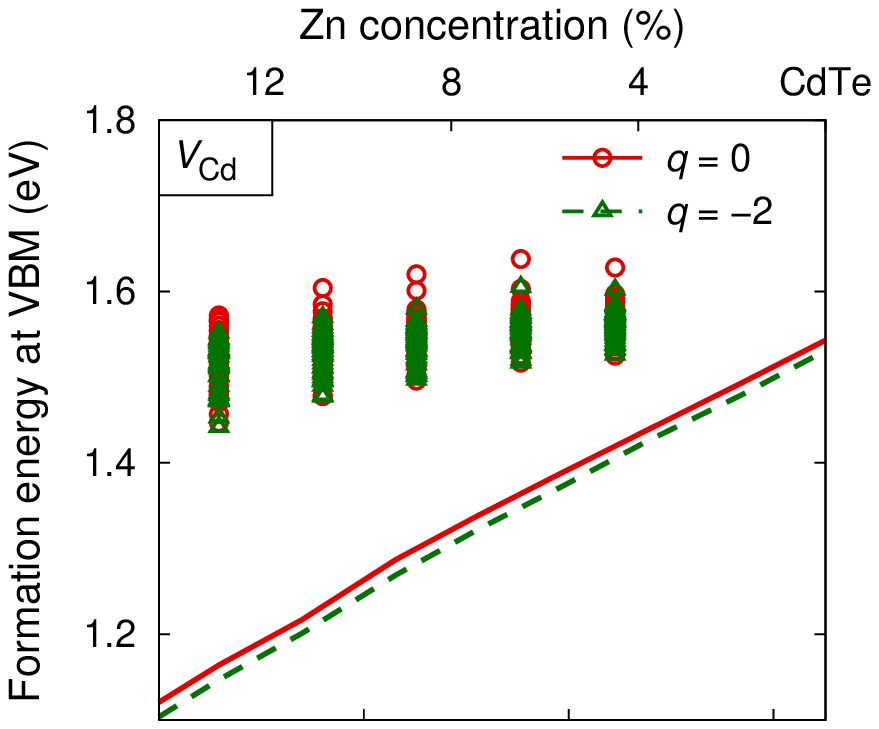} \nolinebreak
 \includegraphics[clip=true,trim=40 40 55 0,width=0.25\paperwidth ]{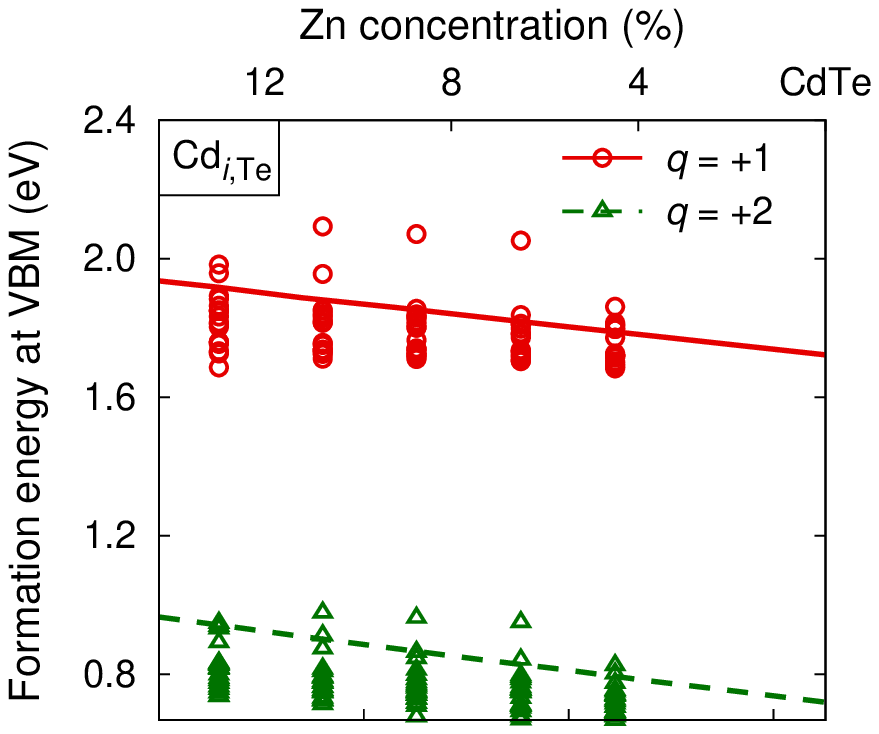} 
 \includegraphics[clip=true,trim=40 0 55 20,width=0.25\paperwidth ]{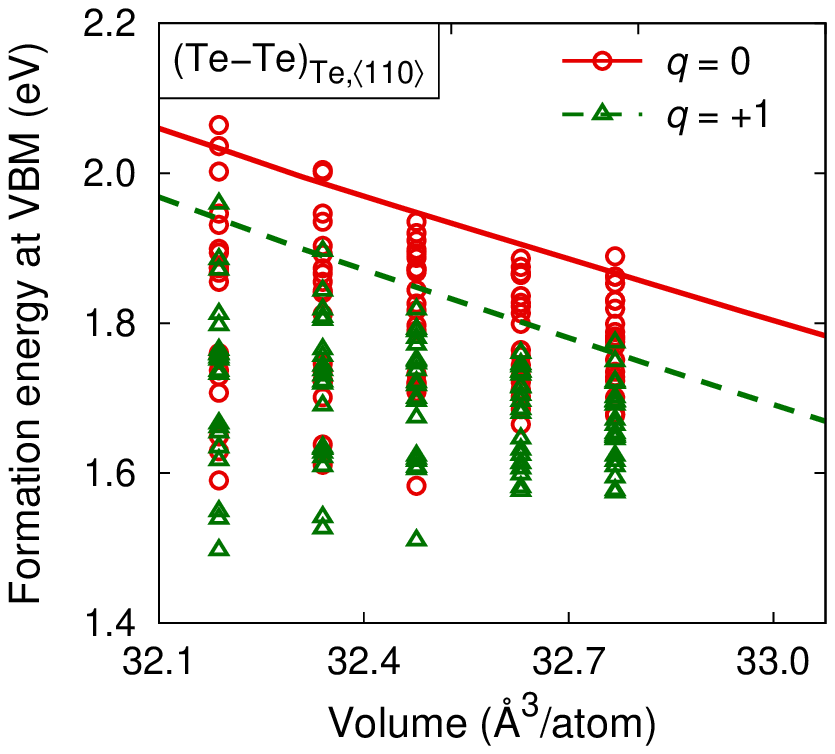} \nolinebreak
 \includegraphics[clip=true,trim=40 0 55 20,width=0.25\paperwidth ]{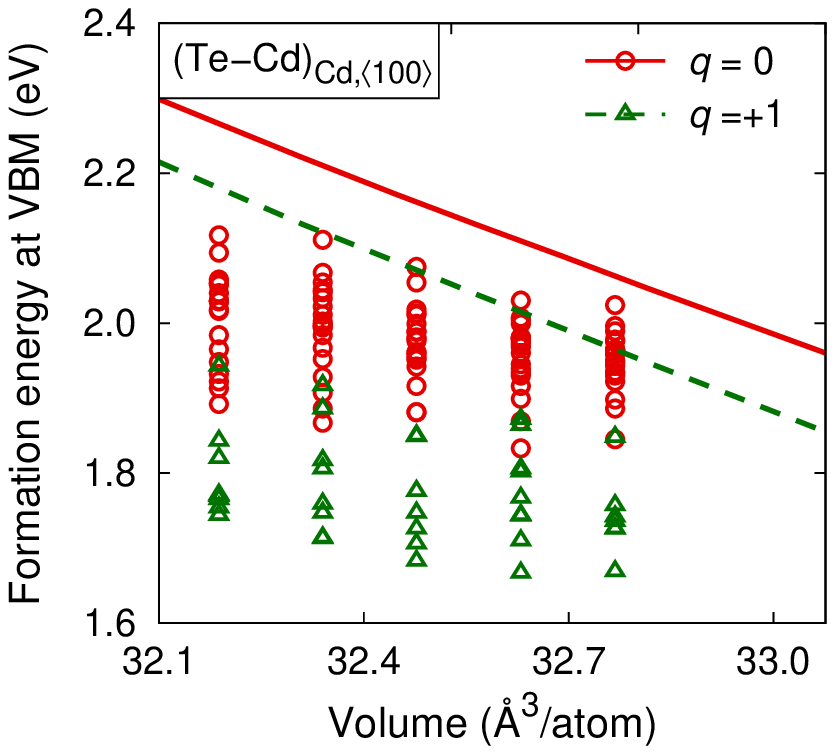} \nolinebreak
 \includegraphics[clip=true,trim=40 0 55 20,width=0.25\paperwidth ]{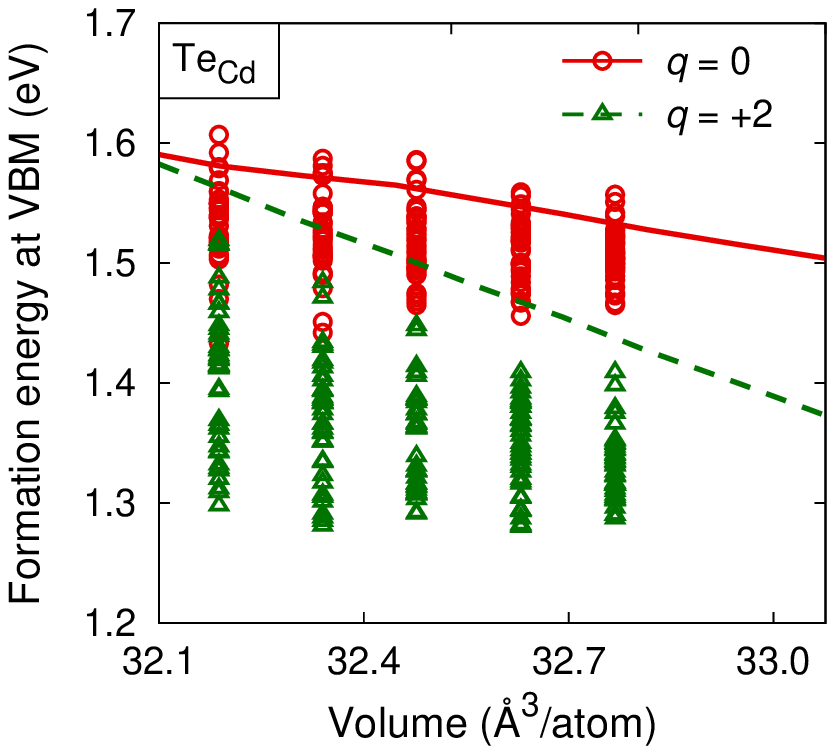}   
  \caption{(Color online) Formation energies at the VBM under Te-rich conditions for the 6 most important intrinsic defects in CdTe as a function of compressive strain (lines) and \CZT\ as a function of volume (symbols).
  }
  \label{fig:eform_strain}
\end{figure*}

The general trends for all defects studied here are clear across the concentration range of interest:  the formation energies increase or stay constant as a function of Zn concentration for all defects and charge states, except $V_\Cd$. Carvalho \textit{et al.} examined the correlation between point defect formation energies for cation-site defects in CZT and volumetrically strained CdTe  and concluded that in the range of $x<0.5$, the volume change induced by alloying is the dominant factor in the change of formation energies.\cite{CarTagObe10} Thus, they posit that accurate defect formation energies for the alloy may be predicted by performing calculations with the pure binary compound using the lattice constant of the alloy. To the benefit of this conclusion, we show in \fig{fig:eform_strain} that usually point defect formation energies in CZT and strained CdTe indeed exhibit some general overall correlation. However, the absolute values of formation energies as well as their variation with volume are quantitatively rather different. In fact, the only defect showing reasonably good quantitative agreement is the Cd interstitial. Hence we conclude that formation energies calculated using the strained binary compound in general have small, if any, predictive power.

\subsection{Optical transition levels} 

\begin{figure}
  \includegraphics[width=\columnwidth]{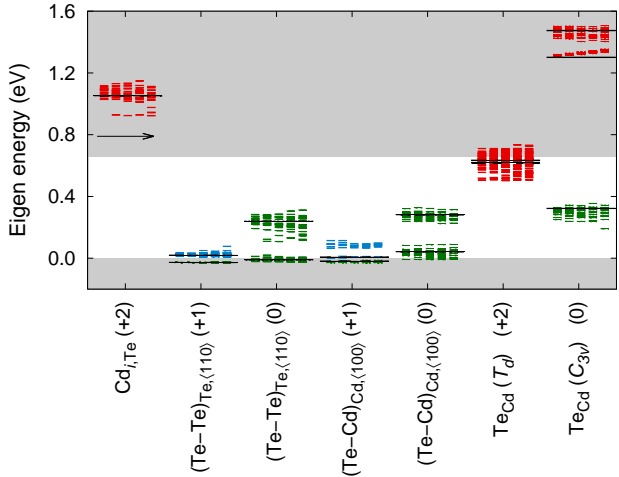}
  \caption{(Color online) Kohn-Sham defect levels for the most important intrinsic defects in \CZT. For each defect, levels of all the randomized cells are shown for five different concentrations ranging from 5.5 to 13.0\,\% (left to right, indicated by arrow). Red, blue, and green line colors denotes an empty, half-filled, and fully occupied level, respectively. The solid black lines show the level positions in the pure CdTe case.
  }
  \label{fig:levels}
\end{figure}

The optical transition levels for defects with levels inside or near the band gap have been extracted from the Kohn-Sham band structure of the defect cell and are displayed in \fig{fig:levels} as a function of Zn concentration. The black solid lines indicate optical transitions in pure CdTe, and each colored line represents a defect state in a particular randomized supercell. In general, the average for each group of levels coincides roughly with the CdTe result. The alloy configurations show various spreads for each defect, arising for the same reasons as the variations in the formation energies described in \sect{sec:eform}.
Furthermore, with a few exceptions (notably, $\Te_\Cd^0$ and ($\Te-$\Te)$_{\Te,<110>}^0$), the position and spread of the transitions are fairly insensitive to Zn concentration. Since no new levels are introduced inside the gap (nor do any existing levels move appreciably deeper into the gap) from Zn alloying, we conclude that the effect of defect-mediated electron-hole recombinations via the Shockley-Read-Hall mechanism\cite{ShoRea52} does not increase upon Zn alloying.

\subsection{Defect, alloy, and phonon scattering}

\begin{figure}
  \includegraphics[width=\columnwidth]{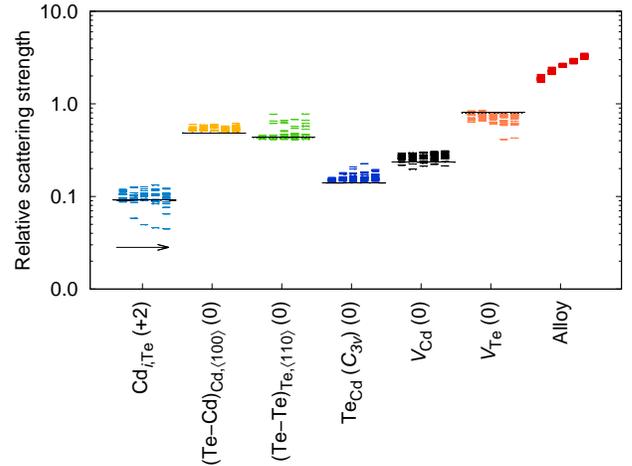}
  \caption{
    (Color online) Relative scattering strengths (Eq. \ref{eq:relscatstr}) for the most important intrinsic defects in \CZT. For each defect, scattering strengths are shown for five different concentrations ranging from 5.5 to 13.0\,\% (left to right, indicated by arrow). The solid black lines show the relative scattering strengths for the pure CdTe case. The last column (Alloy) shows the effect of alloy scattering at the same level of approximation. The data are arbitrarily normalized to the value of the strongest intrinsic defect scattering channel.
  }
  \label{fig:scat}
\end{figure}

The relative carrier scattering strengths for the intrinsic defects as a function of Zn concentration are shown in \fig{fig:scat}.
These scattering strengths correspond to one defect in a 216-atom cell and hence correspond to a defect concentration on the order of $10^{20}$~cm$^{-3}$. Compared to pure CdTe, the scattering strengths are virtually unchanged and exhibit little scatter with alloy configuration. 
A large lattice distortion with respect to the ideal cell gives rise to scattering potentials with large gradients and hence strong scattering. In accordance, the split interstitials and Te vacancies are by far the strongest carrier scattering centers. It is also quite natural to expect that scattering strengths of defects in pure CdTe would serve as a lower bound since the alloy disorder could introduce more complex relaxed geometries as well as stronger scattering potentials. Indeed, this is in fact observed for all defects in the present study with the exception of the neutral Te vacancy. We recall, however, that the nearest neighbors of $V_\Te^0$ are subject to a large inward relaxation, unless it is a Zn atom (see Sect. \ref{sec:eform}). Hence, the defect structure for the alloy is actually {\em less} distorted with respect to the ideal case when a Zn atom sits next to the vacancy, explaining this apparent anomaly.

The last column in \fig{fig:scat} compares the relative scattering of randomized 216-atom cells with respect to the ideal structure at the same lattice constant, representing the effect of alloying. In this case the scattering strength is strongly dependent on the Zn concentration and increases monotonically. As expected, the effect of alloying on carrier scattering  far outweighs defect scattering for any realistic defect concentration.
 
\begin{figure}
  \includegraphics[width=\columnwidth]{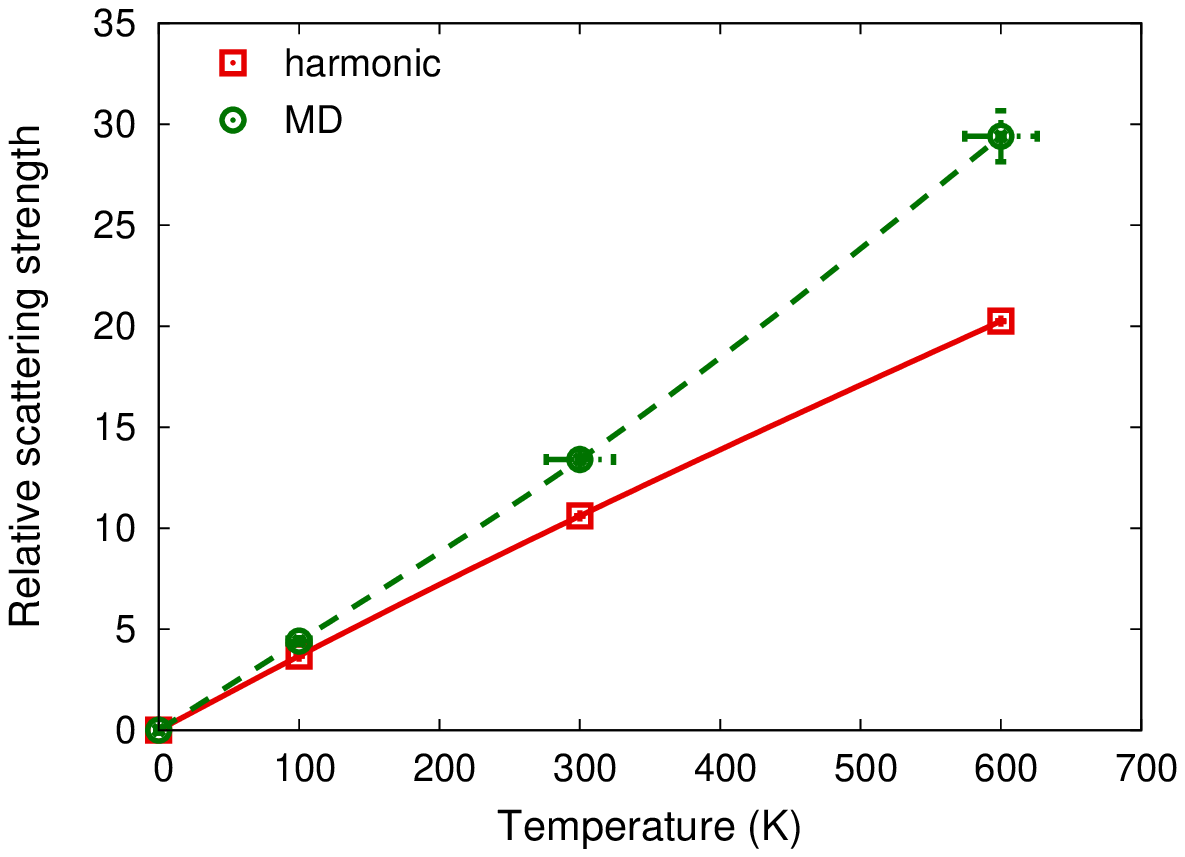}
  \caption{
    (Color online) Relative thermal scattering strengths for CdTe as a function of temperature, calculated using both a harmonic approximation (solid line) and molecular dynamics (dashed line).}
  \label{fig:phonscat}
\end{figure}

The contribution to carrier scattering from phonons is shown in \fig{fig:phonscat}, comparing the results from MD simulations to the harmonic approximation. The latter exhibits a linear temperature dependence whereas the MD data curves upward due to anharmonic effects. The deviation between harmonic and MD data points is a measure for the degree of anharmonicity, which already at room temperature amounts to 20\%\ of the relative scattering strength.
Compared to defect and alloy scattering, these results predict that electronic transport is limited by phonon scattering for temperatures above approximately 150\,K.

\section{Conclusions}
\label{sec:conclusions}
The effects of alloying on pure bulk properties, thermodynamic properties of intrinsic point defects, and carrier scattering rates were examined for CZT alloys with up to 13\%\ Zn. The defect formation energies of the most important intrinsic defects increase or remain constant with increasing Zn content with the exception of the Cd vacancy and to some extent the neutral Te vacancy. The latter defect also deviates from the standard inward relaxation of nearest neighbor atoms in the presence of Zn NNs, creating a very complex and rich energy landscape as a function of Zn coordination (reminiscent of the effects of In coordination in the dilute nitride alloy GaInNAs\cite{LordiGaInNAs1, LordiGaInNAs2}). A previously reported relation between strained CdTe and CZT \cite{CarTagObe10} was revisited, revealing that the dependence of defect formation energies on Zn concentration is at most qualitatively similar to the behavior deduced from volumetrically strained CdTe.

Optical transition levels exhibit small variations with Zn content and introduce no additional defect levels within the gap. Thus Zn alloying should not increase the defect-mediated electron-hole recombinations. 

The  relative carrier scattering rate concept introduced in Ref. \onlinecite{LorErhAbe10} was applied here to alloy and phonon scattering as well as defect scattering. The intrinsic defect scattering in CZT was shown to be very similar to pure CdTe. The scattering rates in CdTe were shown to serve as as lower bound for all defects except the Te vacancy. The latter behavior was explained in terms of local lattice relaxation associated with the distribution of Zn NNs. We furthermore established the relative importance of the different scattering channels and showed that at ambient conditions and moderate defect concentrations the phonon scattering dominates the electric transport properties.

\begin{acknowledgments}
This work was performed under the auspices of the U.S. Department of Energy by Lawrence Livermore National Laboratory under Contract DE-AC52-07NA27344,
with support from the National Nuclear Security Administration Office of Nonproliferation and Verification Research and Development (NA-22).
\end{acknowledgments}
 
%

\end{document}